\documentclass{cargese}  

\let\footnote\savefootnote
\let\footnotetext\savefootnotetext
\let\footnoterule\savefootnoterule
\normallatexbib  

\begin{document}
\articletitle[$AdS_3$ asymptotic (super)symmetries]{$AdS_3$ asymptotic\\
(super)symmetries\footnote{ULB-TH-99/16}}
\author{Karin Bautier}
\affil{Physique Th\'eorique et Math\'ematique \\
Universit\'e Libre de Bruxelles, Campus Plaine C.P. 231 \\
Boulevard du Triomphe, B--1050 Bruxelles, Belgium}

\begin{abstract}
The Chern-Simons formulation of $AdS_3$ supergravity is considered.
A\-symptotic conditions on the Rarita-Schwinger fields are given. Together
with the known boundary conditions on the bosonic fields, these ensure
that the asymptotic algebra is the superconformal algebra, with the same
central charge as the one of pure gravity. It is also indicated that the
asymptotic dynamics is described by
super-Liouville.
\end{abstract}

It has been pointed out in \cite{brown} that the asymptotic
symmetry group of anti-de Sitter gravity in three dimensions is the
conformal group in two dimensions with a central charge $c=3l/2G$. 
This result was obtained by working out explicitly the boundary conditions
and solving the asymptotic Killing equations \cite{brown}. It has
been shown in \cite{vandriel} that the boundary dynamics at infinity is
described by Liouville theory up to terms involving the zero modes and the
holonomies that were not worked out. In the following, we will use the
Chern-Simons formulation of $AdS_3$ (1,1)-supergravity and extend the
analysis of \cite{brown,vandriel} to the supersymmetric case. 
This was originally presented in \cite{banados} written in
collaboration with M. Ba\~nados, O. Coussaert, M. Henneaux and M. Ortiz
and we refer to it for more details.

\paragraph{Chern-Simons action}
$AdS_3$ (1,1)-supergravity can be written as a
Chern-Simons theory \cite{achucarro}. 
The relevant group is $OSp(1|2)\times
OSp(1|2)$ and the action is:
\begin{equation}
I[A,\psi;\tilde{A}, \tilde{\psi}] = I[A,\psi] - I[\tilde{A},
\tilde{\psi}]
\label{a1}  
\end{equation}  
where $I[A, \psi]$ and $I[\tilde{A}, \tilde{\psi}]$
are the Chern-Simons actions for the
supergroup $OSp(1|2)$:
\begin{equation}
I[A,\psi] = \frac{k}{4\pi} \int \left[ \mbox{Tr}(A dA + \frac{2}{3}
A^3)
+ i \bar \psi \wedge D \psi \right]
\label{action}
\end{equation}
(with a similar expression for $I[\tilde{A}, \tilde{\psi}]$). 
The constant $k$ is related to the $3D$ Newton constant $G$ and the
anti-de Sitter radius $l$ through $k = l/4G$.
The action (\ref{action}) is invariant under the bosonic gauge
transformations $\delta_\lambda
A=D\lambda$, $\delta_\lambda\psi=-\lambda\psi$ and under the fermionic
ones $\delta_\rho A^a=i\bar\rho\gamma^a\psi$, $\delta_\rho\psi=D\rho$,
where the gauge parameters $(\lambda^a,\rho) \in osp(1|2)$.

\paragraph{Boundary conditions}
The boundary conditions at infinity on the bosonic fields for an
asymptotically AdS space have been
given in \cite{brown} in the metric representation and they
were reexpressed in the connection representation in \cite{vandriel}. 
To supplement these conditions by appropriate boundary conditions on
the fermionic fields, one follows
the procedure of \cite{teitelboim}: one starts with the known physical
metrics that should be included in the theory - here, the black hole
solutions \cite{BTZ} - and acts on them with the anti-de Sitter
supergroup. 
One then obtains, for one $OSp(1|2)$ copy, the following boundary
conditions:
\begin{eqnarray}
A_v = 0, \ & \ \psi_v = 0, \label{chi} \\
A_r = b^{-1} \partial_r b, \ & \ \psi_r = 0 \label{r}
\end{eqnarray}
and
\begin{equation}
l A_u =  b^{-1} \left(\begin{array}{cc}
                     0  &  L/k \\
                     1  &   0
                   \end{array} \right) b,
\ \ l  \psi_u =
    b^{-1}\left(\begin{array}{c}
                              Q/k  \\
                                 0
                   \end{array} \right)
\label{A+}
\end{equation}
asymptotically.  Here, $u = t +l \phi$, $v = t-l \phi$ and
$L=L(t,\varphi)$ and $Q=Q(t,\varphi)$ are arbitrary functions
which will be shown to be equal to the generators of the
super-Virasoro
algebra.  The group element $b(r)$ is equal to
\begin{equation}
b(r) = \left(\begin{array}{cc}  (r/l)^{1/2} &  0 \\
                                            0  &  (r/l)^{-1/2}
                   \end{array} \right)
\end{equation}
The fact that $\psi_u$ has one zero component implies that it is an
eigenstate of the radial $\gamma$-matrix, which makes the induced
spinors chiral in two dimensions.
The other $OSp(1|2)$ fields satisfy analogous boundary conditions, with
$u$ and $v$ interchanged, and depends on two additional functions $\tilde
L$ and $\tilde Q$. For positive values of $L_0$ and $\tilde{L}_0$, the
boundary conditions (\ref{A+}) represent a black hole, with
$M=(2/k)(L_0+\tilde{L}_0)$ and $J=(2l/k)(\tilde{L}_0-L_0)$. Anti-de Sitter
space corresponds to $L/k=\tilde L/k=-1/4$.

\paragraph{Asymptotic symmetry}
The most general supergauge transformations that preserve the boundary
conditions (\ref{chi}), (\ref{r}) and (\ref{A+}) are characterized by
gauge parameters $(\lambda^a, \rho)$
that must fulfill, to leading order,
\begin{equation}
\lambda(u, r) = b^{-1} \eta(u) b, \ \ \rho(r,u) = b^{-1}
\varepsilon(u)
\label{geneform}
\end{equation}
with
\begin{eqnarray}
\eta^+ &=& \frac{\eta^- L}{k}
 - (1/2) (\eta^-)''+ \frac{iQ\epsilon}{2 k}, \label{p1}  \\
\eta^1 &=& - (\eta^-)',      \label{p2}        \\
\varepsilon &=& \left(\begin{array}{c}  
           -\epsilon' +\eta^- Q/k \\
           \epsilon
                   \end{array} \right),
\label{res-susy}
\end{eqnarray}
where $'$ denotes derivative with respect to the argument,
and $\eta^-$ and $\epsilon$ are two residual functions of $u$.
The asymptotic symmetry acts on the components $L$ and $Q$ of the
connection that remain at infinity in the following way:
\begin{eqnarray}
\delta L &=& (\eta^- L)' + (\eta^-)' L  - \frac{k}{2} (\eta^-)'''
+(\frac{iQ\epsilon}{2})' + i Q \epsilon',
\label{dL} \\
\delta Q &=&  -k\epsilon'' + L \epsilon
+(\eta^- Q)' + \frac{1}{2} (\eta^-)' Q.
\label{dQ}
\end{eqnarray}  
This indicates that $L$ and $Q$ form
a super-Virasoro algebra, with a central charge $c$ equal to
$6k$ ($c/12 = k/2$).

To express this asymptotic symmetry in terms of Poisson brackets,
we note that the canonical generators of the gauge transformations
are:
\begin{equation}
G(\lambda^a)   = \int_\Sigma \lambda^a {\cal G}_a + B , \;
S(\rho) = \int_\Sigma \bar \rho {\cal S} + F. \label{S}
\end{equation}
The boundary terms $B$ and $F$ must be chosen so that the generators
$G$ and $S$ have well-defined functional derivatives \cite{regge}.
Taking into account the above asymptotic conditions and gauge
parameters, one obtains:
\begin{equation}
B = \frac{1}{2\pi}\int_{\partial\Sigma} \eta^- L,
\ \ \ \ F = \frac{-i}{2\pi}\int_{\partial\Sigma}
 \epsilon Q.
\label{surfT}
\end{equation} 
In the reduced phase space (where the constraints are zero), the
generators of the asymptotic symmetry reduce to their boundary term
(\ref{surfT}). From equations (\ref{dL}) and (\ref{dQ}), we compute
their Poisson brackets and we get, after Fourier transformation, the
asymptotic super-Virasoro algebra:
\begin{eqnarray}
~[L_m,L_n] &=& (n-m) L_{n+m} + \frac{k}{2} n^3 \delta_{n+m,0}\\
~[L_m,Q_n] &=& \left(\frac{1}{2}m - n\right) Q_{m+n} \\
~\{Q_m,Q_n\}&=& 2L_{m+n} + 2k  m^2 \delta_{m+n, 0}
\end{eqnarray}
with a central charge equal to $c=6k=3l/2G$. 

\paragraph{Dynamics at infinity}
Following \cite{vandriel} and refering to
\cite{banados} for details and references, we note that
the Chern-Simons theory under the boundary
conditions (\ref{chi}) induces the chiral Wess-Zumino-Witten model at the
boundary. The other boundary conditions (\ref{A+}) turn out to be
precisely the constraints that reduce the WZW theory based on the
supergroup $OSp(1|2)$ to chiral $2D$ supergravity. 
The two chiral theories are combined to get the non chiral
super-Liouville theory (up to zero modes and holonomies). This has been
checked at the level of the action, using the Gauss
decomposition for $OSp(1|2)$.

\begin{acknowledgments}
I woulk like to thank my collaborators M. Ba\~nados, O. Coussaert, M.
Henneaux and M. Ortiz, and the organizers of the Carg\`ese 99 ASI for a
very pleasant school and for financial support. 
This work has been partly supported by the ``Actions de
Recherche Concert{\'e}es" of the ``Direction de la Recherche
Scientifique - Communaut{\'e} Fran{\c c}aise de Belgique" and by
IISN - Belgium (convention 4.4505.86).
The author is ``Chercheur F.R.I.A.'' (Belgium).
\end{acknowledgments}

\end{document}